\input{aipcheck}

\documentclass[
    ,final            
  ]
  {aipproc}

\layoutstyle{6x9}
\usepackage{amsmath}
\usepackage{amssymb}
\usepackage{graphicx}

\begin{document}

\title{Information Theory of Quantum Systems with some hydrogenic applications}


\author{J.S. Dehesa}{
  address={Dpto. de F\'isica At\'omica, Molecular y Nuclear, Universidad de
Granada},altaddress={Instituto Carlos I de F\'{\i}sica Te\'orica y
Computacional,
Universidad de Granada}
}

\author{D. Manzano}{
  address={Dpto. de F\'isica At\'omica, Molecular y Nuclear, Universidad de
Granada},altaddress={Instituto Carlos I de F\'{\i}sica Te\'orica y
Computacional,
Universidad de Granada}
}

\author{P. S\'anchez-Moreno}{
  address={Dpto. de Matem\'atica Aplicada, Universidad de Granada}
  ,altaddress={Instituto Carlos I de F\'{\i}sica Te\'orica y Computacional,
Universidad de Granada} 
}

\author{R.J. Y\'a\~nez}{
  address={Dpto. de Matem\'atica Aplicada, Universidad de Granada}
  ,altaddress={Instituto Carlos I de F\'{\i}sica Te\'orica y Computacional,
Universidad de Granada} 
}

\begin{abstract}
The information-theoretic representation of quantum systems, which complements
the familiar energy description of the  density-functional and
wave-function-based theories, is here discussed. According to it, the internal
disorder of the quantum-mechanical non-relativistic systems can be quantified
by various single (Fisher information, Shannon entropy) and composite (e.g.
Cram\'er-Rao, LMC shape and Fisher-Shannon complexity) functionals of the
Schr\"odinger probability density $\rho(\vec{r})$. First, we examine these
concepts and its application  to quantum systems with central potentials.
Then, we calculate these measures for hydrogenic systems,
emphasizing their predictive power for various physical phenomena. Finally, some
recent open problems are pointed out.
\end{abstract}

\maketitle


\section{Introduction}

The internal disorder of quantum-mechanical non-relativistic systems is
conditioned by the spatial spreading of their Schr\"odinger single-particle
probability density $\rho(\vec{r})$ to a great extent. To quantify it, various
single and composite information-theoretic measures \cite{DEH} have been
proposed beyond the familiar ordinary moments $\left\langle r^k \right\rangle$
of $\rho(\vec{r})$, 
$$
\left\langle r^k \right\rangle := \int_\Omega r^k \rho(\vec{r}) d\vec{r}; \quad
k=0,1,2,\ldots
$$
and the statistical variance
\begin{equation} \label{eq:variance}
V[\rho] = \left\langle r^2 \right\rangle - \left\langle r \right\rangle^2
\end{equation}
Among the single information-theoretic measures it is worth highlighting the
Fisher information \cite{FRI}
\begin{equation}
 F\left[\rho \right]:= \int_\Omega \frac{\left[ \bar{\nabla}\rho (\vec{r})
\right]^2 }{\rho(\vec{r})} d\vec{r}
\end{equation}
and the Shannon entropy \cite{SHA}
\begin{equation}
 S\left[ \rho \right]:=  \int_\Omega \rho (\vec{r})  \log \rho(\vec{r}) d\vec{r}
\end{equation}
and its generalization, the entropic moments
\begin{equation}
W_q[\rho] :=\int_\Omega \left[\rho(\vec{r})\right]^q d\vec{r};\quad q\in
\mathbb{R}
\end{equation}
or its modified forms known as Renyi and Tsallis entropies.

The (translationally invariant) Fisher information $ F\left[\rho \right]$,
contrary to the Shannon, Renyi and Tsallis entropies, is a local measure of
spreading of the density $\rho(\vec{r})$ because it is a gradient functional of
$\rho(\vec{r})$. The higher this quantity is, the more localized is the
density, the smaller is the uncertainty and the higher is the accuracy in
estimating the localization  of the particle. In contrast, the Shannon, Renyi
and Tsallis entropies are global measures of delocalization because they are
logarithmic and power functionals of $\rho(\vec{r})$. They measure the total
extent of the density in various complementary ways, without any reference to
any specific point of the domain of definition of $\rho(\vec{r})$. The latter
should be realized when comparing these three entropies with the variance
$V[\rho]$.

These single information-theoretic measures have been shown to be very fertile
in numerous scientific, technological and financial fields, particularly to
identify, characterize and interpret numerous atomic and molecular phenomena
such as e.g., correlation properties in atoms, spectral avoided crossings of
atoms in external fields and the transition state and the other stationary
points in chemical reactions.

Recently, some information-theoretic measures composed by two of the aforementioned
single measures have been shown to be most appropriate to grasp the different
facets of the internal disorder of quantum systems and to disentangle among
their rich three-dimensional geometries. This is basically because (i) they are
invariant under replication, translation and scaling transformations, and (ii)
they have minimal values for both extreme cases: the completely ordered systems
(e.g. a Dirac delta distribution and a perfect crystal in one and three
dimensions) and the totally disordered systems (e.g., an uniform or highly flat
distribution and an ideal gas in one and three dimensions). The latter property
shows that these two-ingredient complexity measures quantify how easily a
system may be modelled. These composite quantities are the LMC shape complexity
$C_{\text{LMC}}[\rho]$ which is defined \cite{CAT} as
\begin{equation} \label{eq:lmc}
C_{\text{LMC}}[\rho] =D[\rho] \times \exp \left( S[\rho]\right),
\end{equation}
the Fisher-Shannon complexity $C_{\text{FS}}[\rho]$, defined in \cite{AN1} by
\begin{equation} \label{eq:6}
 C_{\text{FS}}[\rho] = F[\rho] \times J[\rho] 
\end{equation}
and the Cram\'er-Rao complexity \cite{AN1}, given by
\begin{equation} \label{eq:cr}
C_{\text{CR}}[\rho] =F[\rho]\times V[\rho]
\end{equation}
where $D[\rho]=W_2[\rho]$ is the second-order entropic moment, heretoforth
called disequilibrium because it quantifies the departure of $\rho(\vec{r})$
from equilibrium, and $J[\rho]$ is the Shannon entropy power defined by
\begin{equation} \label{eq:7b}
 J[\rho]=\frac{1}{2\pi e}\exp
\left(\frac{2}{3} S[\rho]\right)
\end{equation}
for the three-dimensional systems. We note that the LMC shape complexity
measures the combined effect of the average height and the total spreading of
the density, while the Fisher-Shannon complexity grasps the oscillatory nature
of the density together with its total extent in the configuration space, and
the Cram\'er-Rao quantity takes into account the gradient content of the density
jointly with its concentration around the centroid. It is worth noting that
these complexity measures are lower-bounded \cite{LAA} as 
\begin{equation} \label{eq:8b}
C_{\text{LMC}}[\rho]\ge 1,\qquad  C_{\text{FS}}[\rho]\ge 3\qquad \text{and}
\qquad C_{\text{CR}}[\rho]\ge 9
\end{equation}
for general three-dimensional systems.

Here we first survey the information theory of quantum systems subject to
central potentials and then we apply it to hydrogenic systems. Finally some
applications to relativistic and multidimensional systems are shown to
illustrate the predictive power of the theory.

\section{Information Theoretic Measures  of Central Potentials}

The quantum stationary states of a single-particle system in a
spherically-symmetric potential $U(r)$ are known to be described by the wave
functions
\begin{equation}
 \Psi_{nlm} (\vec{r})=R_{nl}(r) Y_{lm} (\theta,\phi)
\end{equation}
with the quantum numbers $n=1,2,\ldots$, $l=0,1,2,\ldots, n-1$, and
$m=-l,-l+1\ldots,l$. The angular part is given by the spherical harmonics
\begin{equation}
Y_{lm}(\theta,\phi)= \frac{1}{\sqrt{2 \pi}} C_{l-m}^{(l+m)} \left( \cos \theta
\right) \left( \sin \theta \right)^m e^{im \phi}
\end{equation}
with $0\le\theta\le\pi$ and $0\le\phi\le 2\pi$, and $C_k^{(\alpha)}(x)$ denotes
the familiar Gegenbauer or ultraspherical polynomials. The radial part
$R_{nl}(r)$ is given by an orthogonal hypergeometric function (i.e., it has the
form $\omega^{1/2}(r) y_n(r)$, where $\{y_n(r)\}$ denotes a system of
polynomials orthogonal with respect to the weight function $\omega(r)$)  or
any other special function of applied mathematics. Then the probability to find
the particle between $\vec{r}$ and  $\vec{r}+d\vec{r}$ is given by
\begin{equation} \label{eq:probdens}
 \rho_{nlm}(\vec{r}) d \vec{r}  = |\Psi_{nlm} \left( \vec{r},t \right) |^2
d\vec{r}={D_{nl}(r)} r^2 dr \times {\Pi_{lm} (\Omega)} d
\Omega,
\end{equation}
where the volume element $d\vec{r}=r^2 dr \sin \theta d\theta d\phi \equiv r^2
dr d\Omega$, and the radial and angular probability densities are given by
\begin{equation}
  {D_{nl}(r)}= |R_{nl}(r)|^2 = \omega_l(r) \left[ y_n(r) \right]^2
\end{equation}
and 
\begin{equation}
{\Pi_{lm} (\Omega)}= |Y_{lm}(\theta,\phi) |^2 = \left[ C_{l-m}^{l+m}
\left( \cos \theta \right) \right]^2 \left[ \sin \theta \right]^{2m}
\end{equation}
respectively. The former density gives the probability per radial interval to
find the particle in $({r},{r}+d{r})$, and the latter one describes  the
spatial profile of the system.

The information theory allows us to calculate the intrinsic randomness
(uncertainty) and the profile of the system by means of various
information-theoretic measures (already defined) of the radial and angular
probability densities, respectively. Beyond the variance, which is given by 
\begin{equation} \label{eq:12}
 V\left[\rho_{nlm} \right]=\int_{0}^{\infty} r^4 \omega_l(r) \left[y_n(r)
\right]^2 dr - \left| \int_{0}^{\infty}  r^3 \omega_l(r) \left[y_n(r) \right]^2
dr \right|^2,
\end{equation}
the Fisher information can be expressed by
\begin{equation} \label{eq:13}
 F \left[\rho_{nlm} \right]=F[y_n]+ \langle r^{-2}
\rangle F[Y_{lm}],
\end{equation}
 where $\langle r^{-2} \rangle$ is equal to the norm of the orthogonal
polynomials $y_n(r)$ involved in the wavefunction, and $F[y_n]$ and $F[Y_{lm}]$
are the Fisher functionals of the polynomial $y_n(r)$ and the spherical
harmonics $Y_{lm}(\Omega)$, respectively. Moreover, the Shannon entropy turns
out to be the sum of the entropic functionals  $E[y_n]$ and $E[Y_{lm}]$, given
by
\begin{equation}
 E[y_n]=-\int_{0}^{\infty} \omega_l(r) y_n^{2}(r) \log \left[ \omega_l(r)
y_n^{2}(r) \right] dr
\end{equation}
and
\begin{equation}
 E[Y_{lm}]=-\int_{0}^{\pi} \sin \theta d \theta \int_{0}^{2 \pi} d\phi
|Y_{lm}(\theta,\phi)|^2 \log |Y_{lm}(\theta,\phi)|^2 
\end{equation}
The corresponding Cram\'er-Rao, Fisher-Shannon and LMC complexities can be
subsequently obtained from their definitions (\ref{eq:lmc})-(\ref{eq:cr}),
respectively, as products of two functionals of orthogonal polynomials of
power, gradient or logarithmic type. 

To go further in the determination of the spreading measures of the system we
need to specify the analytic form of the central potential. This is done for
hydrogenic systems, which have a Coulomb potential, in the following section.

\section{Hydrogenic Information Theory}

For a central potential of Coulombian form, i.e. $U(\vec{r})=-\frac{Z}{r}$, the
quantum stationary states of a hydrogenic system have the energies
$E_n=-\frac{Z^2}{2n^2}$ and the probability density (\ref{eq:probdens})
$\rho_{nlm}(\vec{r})$ given by
$$
\rho_{nlm}(\vec{r})=R_{nl}^2(r) \hspace{1mm} |Y_{lm}(\theta,\phi)|^2
$$
where
$$
R_{nl}^2(r)= \frac{4Z^3}{n^4} {\tilde{r}}^{-1} \omega_{2l+1} (\tilde{r})
\left[{\tilde{L}}_{n-l-1}^{(2l+1)} (\tilde{r})\right]^2
$$
with $\tilde{r}=\frac{2 Z}{n} r$ and $\omega_{\alpha}(\tilde{r})=
\tilde{r}^\alpha e^{-\tilde{r}}$. Taking into  account this fact and realizing
that $y_n(r)$ are the Laguerre polynomials, we can obtain from Eqs.
(\ref{eq:12}) and (\ref{eq:13})  the following values \cite{DLM}
\begin{equation} \label{eq:16}
 V\left[\rho_{nlm} \right]= \frac{1}{4Z^3} [n^2 (n^2+2)-l^2 (l+1)^2]
\end{equation}
and
\begin{equation}\label{eq:17}
 F \left[\rho_{nlm} \right]= \frac{4Z^3}{n^3} (n-|m|) 
\end{equation}
for the variance and Fisher information of any hydrogenic system, respectively.
Note that they depend neither on $m$ (the variance) nor on $l$ (the Fisher
information). Moreover, we can obtain that the Shannon entropy has the
expression
\begin{equation} \label{eq:18}
 S[\rho_{nlm}]=B(n,l,m)-3\log Z
\end{equation}
with
$$
B(n,l,m)=A_1(n,l,m) + \frac{1}{2n}
{E_1\left[L_{n-l-1}^{(2l+1)}\right]} +
{E\left[C_{n-|m|}^{(|m|+\frac{1}{2})}\right]}
$$
being $A_1(n,l,m)$ a known expression, and where the entropic integrals
$$
{E_i[y_k]} := -\frac{1}{\pi} \int_a^b x^i y_k^2(x) \log
\left[y_k^2(x)\right] \omega(x) dx ;\quad i=0,1
$$
cannot be explicitly calculated except for circular $(n,l=m=n-1)$ and Rydberg
$(n\gg 1)$ states. In particular, we obtain the values
$$
S[\rho_{100}] = 3 + \log \pi -3 \log Z 
$$
and
$$
S[\rho_{n00}] = 6 \log n -\log2 + 2 \log \pi+o(1)
$$
for the Shannon entropies of the ground state and the $(ns)$-Rydberg states,
respectively.

Let us now calculate the complexity measures of the hydrogenic system
\cite{DEH1} From Eqs. (\ref{eq:lmc}) and (\ref{eq:18}) jointly with the fact
 the disequilibrium $D[\rho]=Z^3 A_2(n,l,m)$, being $A_2(n,l,m)$ a known
expression, we obtain that the LMC shape complexity has the values
\begin{equation} \label{eq:19}
  C_{\text{LMC}}[\rho_{nlm}]=A_2(n,l,m) e^{B_2(n,l,m)}
\end{equation}
where $A_2$ and $B_2$ are explicitly known functions of the three quantum
numbers $(n,l,m)$. Similarly, from Eqs. (\ref{eq:cr}), (\ref{eq:16}) and
(\ref{eq:17}) we have the following value
\begin{equation}\label{eq:20}
 C_{\text{CR}}[\rho_{nlm}] = \frac{n-|m|}{n^3} \left(n^2 (n^2+2) -
l^2(l+1)\right)^2
\end{equation}
for the Cram\'er-Rao complexity of the hydrogenic system. Moreover, Eqs.
(\ref{eq:6}), (\ref{eq:7b}), (\ref{eq:13}) and (\ref{eq:18}) yield the
following value
\begin{equation}\label{eq:21}
  C_{\text{FS}}[\rho_{nlm}]=\frac{4(n-|m|)}{n^3} \frac{1}{2\pi e} 
e^{\frac{2}{3}B(n,l,m)}
\end{equation}
 for the Fisher-Shannon complexity of the hydrogenic system. 

We observe that the three complexity measures do not depend on the Coulombian
strength, i.e. the nuclear charge $Z$. Moreover, the three complexities,
according to Eqs. (\ref{eq:19}), (\ref{eq:20}) and (\ref{eq:21}), increase with
increasing $n$ for $(l,m)$ fixed, and they decrease with increasing $l$ for
$(n,m)$ fixed; the latter behaviour is also observed with $m$ for $(n,l)$
fixed. Finally, for completeness, we have plotted in Figure 1 the variation of
the Fisher-Shannon complexity (\ref{eq:18}) in terms of $n$ for various
quasi-circular $(l=n-1)$ hydrogenic states. Therein, we observe the increasing
value of the measure when the principal quantum number is increasing, clearly
illustrating the idea of complexity in everyday life. Finally, let us also
collect the values
$$
C_{\text{CR}}[\rho_{100}] = 3,\qquad C_{\text{LMC}}[\rho_{100}] = \frac{e^3}{8}
\quad \text{and} \quad C_{\text{FS}}[\rho_{100}] =\frac{2e}{\pi^{1/3}}
$$
for the three complexity measures of the hydrogenic ground state.

\begin{figure}
  \includegraphics[height=5cm]{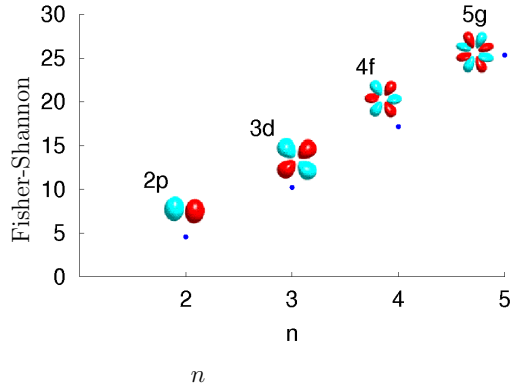}
  \caption{Fisher-Shannon complexity for quasi
circular $(l=n-1)$ hydrogenic states in terms of the principal quantum number $n$.}
\end{figure}

\section{Relativistic and Dimensionality Applications}

Here we briefly consider some extensions of the previous information theory of
hydrogenic systems to illustrate the predictive  power of the complexity
measures for relativistic and dimensionality phenomena. First, we show that the
Fisher-Shannon complexity is a good quantitative indicator of the relativistic
effects of hydrogenic systems in the Klein-Gordon framework. This is done by
calculating this measure for pionic atoms with nuclear charge Z \cite{MAN},
taking care of the Lorentz invariance. In this case, the charge density has the
expression
\begin{equation} \label{eq:24}
\rho(\vec{r}) = \frac{e}{m_o c^2}\left[ \epsilon - \frac{Ze^2}{r}\right]
\left|\Psi_{nlm}(\vec{r})\right|^2
\end{equation}
where the pionic wavefunction $\Psi_{nlm}(\vec{r})$ is the physical solution of
the Klein-Gordon equation
$$
\left(\epsilon- \frac{Ze^2}{r}\right) \Psi(\vec{r}) = \left( - \hbar c^2
\vec{\nabla}^2 + m_0^2 c^4\right) \Psi(\vec{r})
$$
The Fisher-Shannon complexity (\ref{eq:6}) of the Lorentz-invariant density
given by (\ref{eq:24}) and the duly normalized to the electronic charge $e$,
has been calculated as a function of $Z$. The results are plotted in Figure 2,
where a comparison is also made with the Schr\"odinger or non-relativistic
values of this measure. Therein, it is clear that the Klein-Gordon values of
the Fisher-Shannon complexity do depend on the nuclear charge, contrary to the
Schr\"odinger case (see also Eq. (\ref{eq:21})). This is a clear indication that
this complexity measure allows for a quantitative estimation of the
relativistic effects on the internal disorder of the hydrogenic systems of
pionic type. See Ref. \cite{MAN} for further details.

\begin{figure}
  \includegraphics[height=5cm]{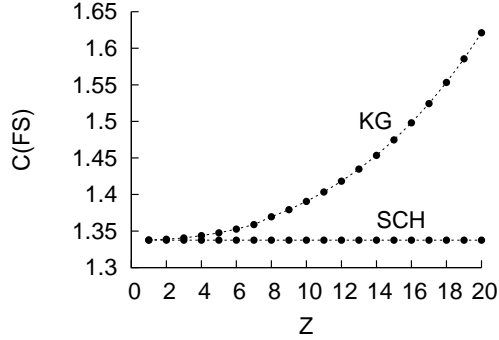}
  \caption{Fisher-Shannon complexity of the ground state of Klein Gordon and
Schr\"odinger Coulomb equation in terms of the nuclear charge $Z$}
\end{figure}

Now, let us consider the LMC shape complexity (\ref{eq:lmc}) of the probability
density
$$
\rho_{\text{C.S.}}(\vec{r}) = \frac{2^{D+2-2n} Z^D \prod_{j=1}^{D-2} \left(\sin
\theta_j\right)^{2n-2}}{\pi^{\frac{D-1}{2}} (2n+D-3)^D \Gamma(n)
\Gamma\left(n+\frac{D-1}{2}\right)} e^{-\frac{r}{\lambda}}
\left(\frac{r}{\lambda}\right)^{2n-2}
$$
corresponding to the circular states (i.e., $l=m=n-1$) of $D$-dimensional
hydrogenic systems \cite{DLM, LOP}, where
$\vec{r}=(r,\theta_1,\theta_2,\ldots,\theta_{D-1})$, $\lambda=\frac{\eta}{2Z}$,
and $\eta=n+\frac{D-3}{2}$ is the principal hyperquantum number. The value has been
recently obtained \cite{LOP}
\begin{multline*}
C_{\text{LMC}}\left[\rho_{\text{C.S.}}\right] =
\frac{\Gamma\left(n-\frac12\right)
\Gamma\left(2n+\frac{D-3}{2}\right)}{2^{2n+D-2} \pi^{1/2}
\Gamma\left(n+\frac{D-1}{2}\right)} \\  \exp\left\{ 2n+D-2-(n-1)
\left[\psi(n)+\psi\left(n+\frac{D-1}{2}\right)\right]\right\},
\end{multline*}
(where $\psi(x)=\Gamma'(x)/\Gamma(x)$ is the digamma function) for the position
LMC complexity of a $D$-dimensional hydrogenic system in an arbitrary circular
state. The corresponding dimensionality dependence of this measure is plotted in
Figure 3 for the ground state $(n=1)$ and the excited circular states with $n=2$
and $3$. Therein, we note a parabolic growth for all states when the
dimensionality is increasing, being always greater than unity in accordance 
with Eq. (\ref{eq:8b}). In addition, we observe that the minimum value of this
complexity measure is $(e/2)^2=1.847$, occurring for $D=2$.

\begin{figure}
  \includegraphics[height=4.7cm]{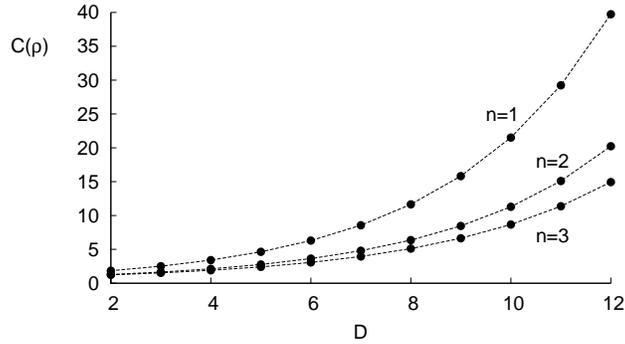}
  \caption{LMC complexity of hydrogenic circular states with $n=1,2$ and $3$ in
terms of the dimension $D$}
\end{figure}

\section{Conclusion and Open Problems}

In summary we have surveyed various aspects of the emerging information theory
of quantum systems with central potentials, emphasizing the results obtained
for hydrogenic states. Furthermore, we have shown the predictive power of the
complexity measures for the relativistic and dimensionality effects of some
ground and excited hydrogenic states.

Let us finally mention a few open problems in this field. First, to study the
analytical and asymptotical properties of the entropic moments $W_q[\rho]$ for
both central and Coulombian potentials. Second, to mutually compare the various
direct spreading measures in the sense of Hall \cite{HAL} for hydrogenic
states. Third, to analyze the behaviour of the complexity measures of
$D$-dimensional systems in the presence of external magnetic and electric
fields. Fourth, to examine the information-theoretic measures of many-electron
atoms moving on a $D$-dimensional hypersphere.

\begin{theacknowledgments}
 We are very grateful for the partial financial support of the grants
FIS2008-02380, FQM-2445, FQM-4643 and FQM-207 of the Junta de Andalucia and
Ministerio de Ciencia e Innovacion, Spain, EU.
\end{theacknowledgments}

\bibliographystyle{aipproc}

\end{document}